# The cosmochemistry of planetary systems

Martin Bizzarro[1,2], Anders Johansen[1,3] & Caroline Dorn[,4]

[1]Centre for Star and Planet Formation, Globe Institute, University of Copenhagen, Copenhagen, Denmark.
[2]Institut de Physique du Globe de Paris, Université de Paris, Paris, France.
[3]Lund Observatory, Department of Astronomy and Theoretical Physics, Lund University, Lund, Sweden.
[4]ETH Zurich, Institute for Particle Physics and Astrophysics, Zurich, Switzerland

**Abstract**

Planets form and obtain their compositions from the leftover material present in protoplanetary disks of dust and gas surrounding young stars. The chemical make-up of a disk influences every aspect of planetary composition including their overall chemical properties, volatile content, atmospheric composition, and potential for habitability. This Review discusses our knowledge of the chemical and isotopic composition of Solar System materials and how this information can be used to place constraints on the formation pathways of terrestrial planets. We conclude that planetesimal formation by the streaming instability followed by rapid accretion of drifting pebbles within the protoplanetary disk lifetime reproduces most of the chemical and isotopic observables in Solar System. This finding has important implications for planetary habitability beyond the Solar System because in pebble accretion, volatiles important for life are accreted during the main growth phase of rocky planets as opposed to the late-stage. Finally, we explore how bulk chemical inventories and masses of planetary bodies control the composition of their primordial atmospheres and their potential to develop habitable conditions.

**[H1] Introduction**

Sun-like stars are born from the gravitational collapse of the dense cores of molecular clouds comprised of dust and gas[1,2]. Different generations of solid material are present in these dense cores, including ancient, galactically-inherited dust, newly-formed supernovae dust from exploding dying stars, interstellar ices, and organic species. Shortly after their formation, Sun-like stars are surrounded by a short-lived protoplanetary disk consisting of gas and solid material inherited from the molecular cloud[3]. Planets are formed and obtain their composition from the material present in protoplanetary disks[4]. Thus, the chemical composition of these planet-forming disks controls all aspects of planetary compositions, including bulk elemental inventories, access to water and other volatiles as well as the complex organic chemistry that is critical to life and, as such, planetary habitability.

The discovery of more than 5,000 exoplanets makes it clear that planets orbiting distant stars are common rather than being rare exceptions, with some estimating that nearly every star in our Galaxy hosts at least one planet[5]. This realization has transformed our understanding of the rate of occurrence of planets and planetary systems and raises the possibility that potentially habitable worlds akin to the Solar System's terrestrial planets may be widespread. Thus, a grand question in planetary sciences is how the formation pathway of rocky planets may influence their chemical make-up and, hence, their potential habitability. This, in turn, requires a full understanding of the mechanisms and timescales of rocky planet formation and evolution.

The Solar System provides a starting point to study rocky planet formation processes because of the wealth of astromaterials available for detailed investigations[6]. In addition to terrestrial samples, meteorites from the surface of the Moon and Mars as well as meteorites from primitive and fully or partially differentiated asteroids allow us to compare the chemical and isotopic make-up of the primordial protoplanetary disk material with that of asteroids and fully grown planets. A key cosmochemical observation is the existence of a gradient in the nucleosynthetic isotope composition of planetary materials with orbital distance, which is interpreted to reflect a primordial protoplanetary disk heterogeneity[7–9]. Material accreted relatively close to the proto-Sun like Earth, Mars, and the parent asteroids of non-carbonaceous (NC) meteorites has a different nucleosynthetic make-up relative to water-rich carbonaceous (CC) asteroids formed in the outer Solar System[10]. Thus, by using nucleosynthetic fingerprinting, it is possible to quantify the relative contribution of inner and outer disk material during the formation of rocky planets and, hence, identify the origin of their precursor material. The classical model of terrestrial planet formation involves a period of giant impacts between embryos over timescales of >100 million years (Myr)[11]. This long-standing idea has been challenged by astrophysical observations[12] and isotopic evidence for rapid planetary accretion[13]. As such, the mechanisms by which terrestrial planets grow are debated and theories such as pebble accretion allowing rapid formation timescales have emerged[14,15]. In pebble accretion, ~100-km-sized bodies first formed by the streaming instability[16] — a hydrodynamical phenomenon that extracts energy from the streaming flow of solids through gas[17]. These bodies then rapidly grow to form terrestrial planets by accretion of millimetre-sized pebbles within the few Myr protoplanetary disk lifetime[18,19]. Although these planet formation mechanisms are not mutually exclusive, they make contrasting prediction with respect to the nature of the terrestrial planets' building blocks. In the planetesimal-driven model, the planetary building blocks are locally derived and dominated by inner Solar System material[20]. In contrast, pebble accretion predicts that terrestrial planets accreted inwardly drifting pebbles from the outer Solar System, such that their final compositions represent a mixture of inner and outer disk material[21]. Clearly, accurate knowledge of the building blocks of the Solar System's terrestrial planets is critical to elucidate their formation pathways and constrain the source of their volatiles, including water.

In this work, we review the currently available cosmochemical data for Solar System solids and asteroids formed during the lifetime of the protoplanetary disk, including their chronology. We compare these data with the chemical and nucleosynthetic make-up of Earth and Mars and discuss the implications for our understanding of planet formation pathways. Both the traditional collisional accretion model of terrestrial planet formation and the pebble accretion model are discussed. We highlight the role of volatility-driven processes during the main accretion phase of protoplanetary bodies as an important driver of chemical and isotopic variability. In the latter sections, we focus on planetary differentiation processes and the partitioning of volatile species between major planetary geochemical reservoirs such as core, mantle, crust, and atmosphere. These results are discussed in the framework of the composition of exoplanetary systems. Finally, we explore how the bulk chemical inventories and masses of planets controls the composition of their primordial atmospheres and, hence, their potential to develop habitable conditions.

## Sampling the Solar System in space and time

The study of meteorites and other extraterrestrial materials provides a unique opportunity to explore the history of our Solar System. These materials serve as time capsules, preserving information about the conditions during the formation of planets, asteroids, and other celestial bodies. Understanding their composition and origin offers valuable insights into the processes that shaped our Solar System.

### *Material available for study*

Meteorites provide by far the greatest source of extraterrestrial material available to us for study. They represent sizeable fragments of asteroids, as well as representing the surfaces of larger bodies like Mars and the Moon that crash-land on Earth. The bulk of the meteorites in our collections come from asteroids located between Mars and Jupiter — in the so-called asteroid belt[22]. Ranging in size from a few tens of meters to more than 500 km in diameter, most asteroids were implanted in the belt during and after the epoch of formation of the inner rocky planets and outer gas- and ice-giants[23,24]. Thus, the asteroid belt is populated by bodies accreted across the Solar System, providing samples of asteroids that formed in the same region as the terrestrial planets, as well as primitive asteroids from the distant and colder outer Solar System.

Asteroids can be broadly grouped into two types, namely non-differentiated and differentiated. Meteorites from these asteroids are dubbed chondrites[25] and achondrites[26], respectively. Differentiated asteroids were once fully molten and crystallized a metallic core, with a silicate mantle and crust. Achondrites represent samples of the different layers of asteroidal bodies, including the core (that is, iron meteorites), the upper mantle (such as, diogenites and ureilites), crust (such as, eucrites and angrites) and even possibly the core and mantle boundary (that is, pallasites). These meteorites provide information about the timescale and formation mechanism(s) of protoplanets. In contrast, non-differentiated asteroids were never molten and, as such, have remained essentially unchanged since their formation. They contain the raw ingredients present in the solar protoplanetary disk at the time when the planets initiated their assembly and allow us to determine the nature and history of the original material from which planets formed. Moreover, rare mm-sized fragments of cometary origin may be preserved in some primitive meteorites[27,28]. An additional source of extraterrestrial material comes from microscopic samples such as interplanetary dust particles[29] (IDPs) and micrometeorites[30]. Ranging in size from tens to several hundreds of microns, these objects represent a dust source from colliding asteroids and possibly also distant comets.

Although meteorites and related extraterrestrial materials provide an invaluable source of information, the bodies they come from are often not known. This knowledge gap can be avoided by directly retrieving samples from other worlds and bringing them back to Earth, a process known as sample return. Efforts like the Hayabusa2 and OSIRIS-Rex space missions[29,30] have focussed on returning samples from the Ryugu and Bennu asteroids, respectively — primitive asteroids believed to have been originally formed in the outer Solar System. Analysis of this highly pristine material allows direct comparison with meteorites thought to come from outer Solar System asteroids, thereby improving our understanding of their origins.

*Formation timescales of solids, asteroids, and protoplanets*

The most abundant primary disk solids preserved in chondrites meteorites are mm- to cm-sized refractory inclusions and chondrules. Refractory inclusions, which comprise calcium-aluminium-rich inclusions (CAIs) and refractory olivine inclusions (that is amoeboid olivine aggregates, AOAs), are thought to have initially formed as gas condensates near the proto-Sun during its earliest evolutionary stage[31]. In contrast, chondrules formed by rapid flash heating and subsequent cooling of pre-existing dusty aggregates in various regions of the protoplanetary disk[32]. The formation timescale of these objects can be deduced by radiometric age dating techniques (Box 1). Focusing on the absolute U-corrected Pb-Pb dating method, the primary formation of refractory inclusions is inferred to have occurred at 4,567.3±0.16 million years ago[33], at an age that is also used to define the birth of the Sun. Chondrule formation, on the other hand, started contemporaneously with refractory inclusions and lasted several million years[33,34]. However, based on Pb isotopes, it is inferred that primary chondrule production was restricted to the first million years after the formation of the Sun and that pre-existing chondrules were recycled and remolten during the remaining lifetime of the protoplanetary disk[34].

Although the parent asteroids of chondrite meteorites have long been considered to represent the precursors of differentiated asteroids, it is now well established that differentiated asteroids were the first bodies to form in the early Solar System[35,36]. This is because the only heat source capable of efficiently inducing asteroidal melting is the short-lived $^{26}$Al radionuclide[37]. Thermal modelling indicate that accretion of differentiated asteroids must have occurred ~0.1 to ~1 Myr after the formation of the first Solar System solids, depending on the assumed initial abundance of $^{26}$Al in the accretion regions of the various bodies[35,38–41]. In contrast, protracted accretion of >2 Myr after Solar System formation is required for the parent asteroids of chondrite meteorites, an epoch when the initial abundance of $^{26}$Al was low enough to avoid considerable heating[42,43]. This apparent bimodality in the accretion ages of differentiated and non-differentiated asteroids (<1 and >2 Myr, respectively) can be understood in the framework of the streaming instability theory of asteroid formation[16]. The high dust density of the early phase of protoplanetary disk evolution (<1 Myr) is favourable for triggering streaming instability[44,45], which can account for efficient and early accretion of the parent asteroids of achondrites. As the protoplanetary disk evolves and the gas is photoevaporated, an increase in dust density occurs, thereby promoting a second phase of asteroidal accretion by streaming instability[46]. Thus, the formation of non-differentiated asteroids may be associated with inside-out gas dissipation at the end of the disk lifetime. Finally, several lines of evidence indicate that Mars, a ~6,800 km diameter planet that can be considered as a remnant protoplanetary body, must also have grown to its size within the ~5 Myr lifetime of the protoplanetary disk. Estimates based on either the timing of core formation[47], magma ocean solidification[48], or primordial crust extraction[49] all support such rapid accretion timescales, which has been linked to the mechanism of pebble accretion[18]. A summary of the current understanding of the chronology of early Solar System events is presented in Fig. 1.

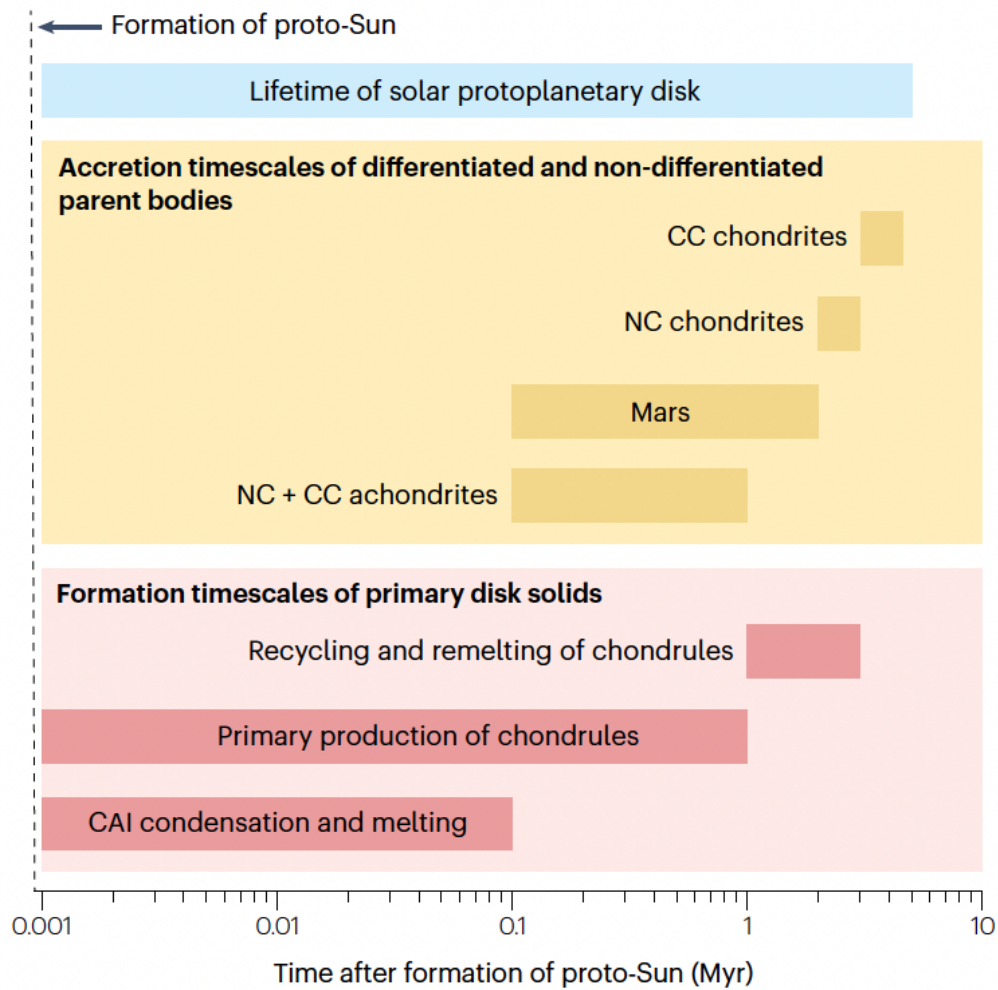

**Fig. 1 | Summary of the chronology of solid formation and accretion of differentiated and non-differentiated bodies.** Ages are expressed relative to calcium-aluminium-rich inclusion (CAI) condensation at 4,567.3±0.16 Myr[33]. Formation timescales of primary disk solids, namely CAIs and chondrules, are based on the U-corrected Pb-Pb system[33,34]. Accretion timescales of the various non-carbonaceous (NC) and carbonaceous (CC) bodies are derived from thermal modelling, assuming that $^{26}$Al was the main heat-source in the early Solar System[35–43] (see Box 1). The lifetime of the protoplanetary disk is constrained by the youngest chondrule age obtained for the Gujba CB$_b$ chondrite[103].

*Volatile depletion of rocky planets*

The bulk chemical composition of the Solar System is essentially that of the Sun, which is made-up of H (74%), He (24%), and a small fraction (2%) of heavier elements, including rock-forming elements such as, for example, O, Si, Al and Fe. The abundance of heavy elements in the Sun provides a fossil record of the initial chemical composition of the protoplanetary disk solids and, by extension, that of the material available to fuel the growth of rocky bodies. Of all the primitive meteorites, only the CI (Ivuna-type) carbonaceous chondrites have elemental abundances that match that of the solar photosphere for heavy elements[50]. Their chemical abundance is therefore used as a proxy for the composition of the collapsing molecular cloud material accreting to the protoplanetary disk.

In contrast to CI chondrites, all remaining chondrite groups, including both non-carbonaceous and carbonaceous chondrites, contain a mixture of low- and high-temperature components[25]. This indicates that an important fraction of their precursors were thermally processed, which is reflected in their chemical composition that is depleted in volatile and moderately volatile elements relative to CI chondrites. Due to viscous heating (the process by which friction and turbulence within the disk convert kinetic energy into heat) and stellar irradiation, accreting protoplanetary disks develop a high temperature gradient outward from the protostar[51]. This high temperature can vaporize existing materials in the inner disk, leading to re-condensation in a sequence recorded by some chondrite minerals[52]. Moreover, shock-related transient heating events provide additional local heat sources to thermally process and devolatilize solids across the protoplanetary disk[53,54]. Thus, multiple processes can lead to efficient volatile depletion of early-formed solids during the entire protoplanetary disk lifetime.

The depletion of volatile and moderately volatile elements observed in most chondrite meteorites is also mirrored on Earth, Mars, and the Moon[55–59] (Fig. 2). In addition to the initial volatile depletion phase associated with the formation of primary disk solids[60], a second phase of volatile loss can occur by planetary processes. Asteroidal differentiation and the development of global magma oceans can effectively lead to evaporative loss of volatile and moderately volatile elements[61]. This process is likely linked to the escape velocity and, hence, more efficient in small airless bodies such as the Vesta asteroid. Thus, in a model where terrestrial planets grow by collisional accretion of smaller bodies, the volatile depletion signature may be inherited from the planetary building blocks. Moreover, additional volatile depletion might have occurred during high-energy impacts associated with the late accretionary stages of planetary bodies such as, for example, the giant impact between Theia and proto-Earth — an additional rocky planet located between the proto-Earth and Mars — that formed the Earth-Moon system[62]. In the pebble accretion model, a growing planet attracts a hydrostatic gaseous envelope that can heat up to several thousands of Kelvin, such that accreting pebbles can be devolatilized during passage through the envelope[63–65]. Thus, volatile depletion appears to be a generic outcome of rocky planet formation, irrespective of the planet-forming mechanism. However, the main distinction in the two planet formation mechanisms is the timing of volatile accretion. In the planetesimal-driven model, the volatile inventory is accreted late by stochastic processes, whereas in pebble accretion, volatiles are accreted early before the establishment of a hot hydrostatic gaseous envelope.

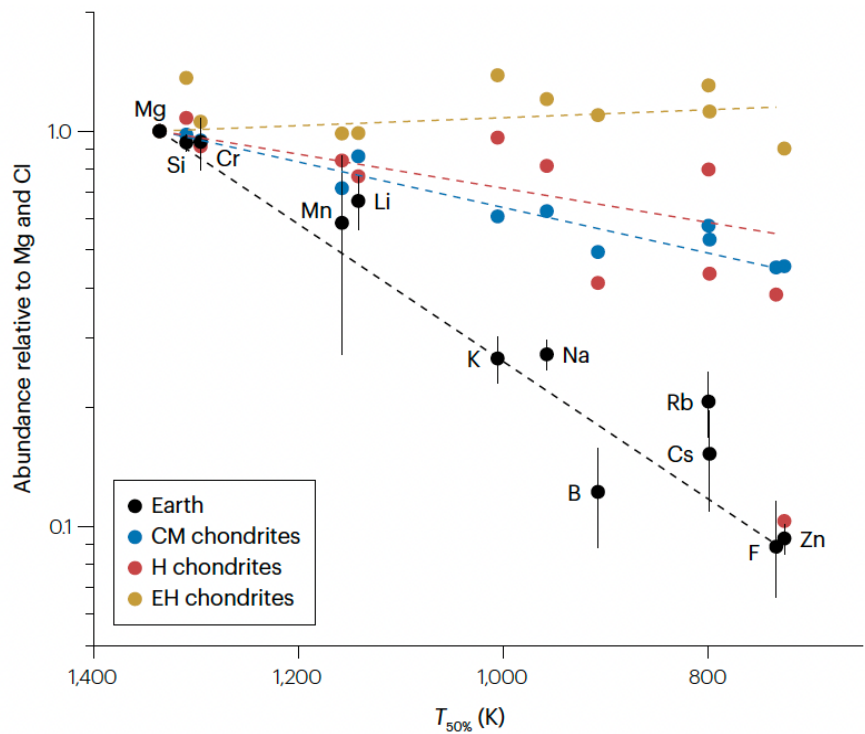

**Fig. 2 | Volatile element depletion patterns.** The abundances of the major lithophile elements more volatile than Mg — relative to Mg and Ivuna-type chondrites (CI) — are plotted against their 50% condensation temperature[174]. Bulk silicate Earth abundances are from ref.[217] and CM, H and EH chondrites are from ref.[218]. Enstatite chondrites (represented here by the EH group) that likely formed in the inner Solar System are surprisingly close to CI chondrites in composition. CM and H chondrites (chosen here to represent the Mighei-type carbonaceous chondrites and ordinary chondrites, respectively) display moderate depletion of volatile elements, but are not nearly as strongly depleted as Earth. The emergence of Earth's volatile depletion pattern requires thermal processing in excess of that experienced by the chondrites.

# The nucleosynthetic variability of the Solar System

A fundamental observation emerging from the analysis of the nucleosynthetic composition of meteorites (Box 2) is that bodies interpreted to have accreted in the volatile-poor inner Solar System have a distinct nucleosynthetic composition relative to volatile-rich bodies formed at larger orbital distances (Fig. 3).

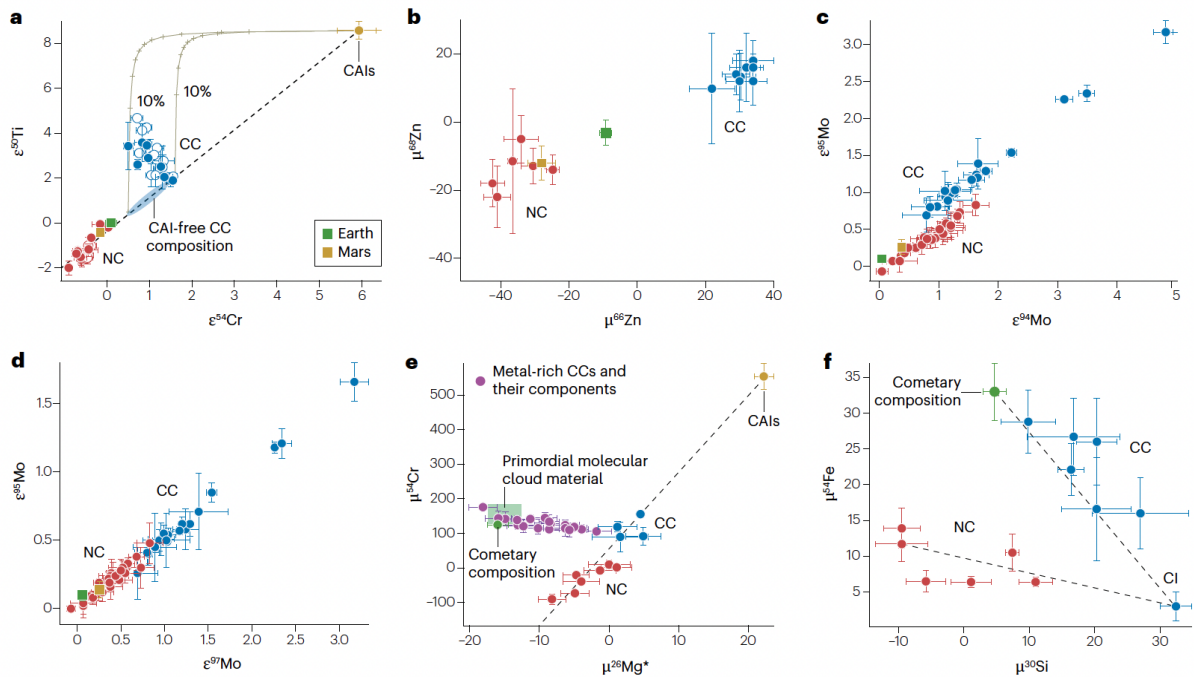

**Fig. 3 | Selected nucleosynthetic tracers. A-F.** The deviations in the isotopic composition of elements relative to that of Earth in parts per million (µ) or parts per 10,000 (ε); $\varepsilon^{50}$Ti versus $\varepsilon^{54}$Cr (**A**), $\mu^{68}$Zn versus $\mu^{66}$Zn (**B**), $\varepsilon^{95}$Mo versus $\varepsilon^{94}$Mo (**C**), $\varepsilon^{95}$Mo versus $\varepsilon^{97}$Mo (**D**), $\mu^{54}$Cr versus $\mu^{26}$Mg* (**E**) and $\mu^{54}$Fe versus $\mu^{30}$Si (**F**). Mixing lines in **A** reflect admixing of CAI material to either a CI composition or a CAI-free hypothetical composition defined by a linear array between, CAIs, NC and CI. Endmember compositions are from refs.[8,219]. These models show that a decoupling in $\varepsilon^{54}$Cr and $\varepsilon^{50}$Ti in CCs can be explained by the addition of small quantities of CAI material (<10 wt %). In **E** and **F**, three endmember compositions are required to explain the variability for the $\mu^{54}$Cr, $\mu^{26}$Mg*, $\mu^{54}$Fe and $\mu^{30}$Si tracers, namely NC, CC and the cometary composition. CAI: calcium-aluminium-rich inclusion; NC: non-carbonaceous chondrites; CC carbonaceous chondrites. Graphs plotted with data from refs.[7,8,28,69,70,173,219–226].

First identified for the neutron-rich iron-group isotope $^{54}$Cr, Trinquier *et al.*[7] demonstrated that carbonaceous chondrites contain excesses of $^{54}$Cr relative to Earth whereas most non-carbonaceous chondrites and achondrites recorded deficits. This pattern also exists for other neutron-rich iron-group element isotopes such as $^{50}$Ti, $^{48}$Ca, $^{64}$Ni (refs. [8,21,66–68]) and for the $^{68}$Zn and $^{70}$Zn nuclides[69–71]. A similar isotopic contrast between NC and CC bodies has also been recognised for heavier elements that comprise isotope species formed by *s*–process, *p*–process, and *r*–process nucleosynthesis, including Zr, Mo, Pd, and Nd (refs. [20,72–75]). Collectively, these data demonstrate that the reservoir from which CC bodies originated is enriched in neutron-rich iron-group element isotopes as well as isotopes formed by the *r*– and *p*–processes relative to the NC reservoir. Another important observation is that CAIs, the first Solar System solids, are typified by even greater enrichments in neutron-rich iron-group element isotopes and *r*– and *p*–process isotopes, than that observed for CC reservoir materials[76,77]. As CAIs originally formed as gas condensates within a few 100,000 years after the formation of the proto-Sun[31], these data suggest the existence of a transient, isotopically anomalous gas reservoir close to the young Sun at early times.

The nucleosynthetic contrast between NC and CC bodies has been interpreted as a compositional dichotomy that requires the spatial isolation of inner and outer disk reservoirs to prevent mixing of the two compositions[10]. Central to this interpretation is the molybdenum isotope composition of NC and CC meteorites[78–80]. In detail, the NC and CC group define distinct trends in an $\varepsilon^{95}$Mo versus $\varepsilon^{94}$Mo diagram (Fig. 3C). Note that the $^{94}$Mo nuclide is essentially a *p*–process isotope, whereas $^{95}$Mo is synthesized by the *s*– and *r*–process. When plotting $\varepsilon^{95}$Mo versus $\varepsilon^{97}$Mo, which have similar nucleosynthesis, this dichotomy is not observed (Fig. 3D, see Box 2 for a definition of $\varepsilon$). Possible mechanisms for the early isolation of inner and outer disk reservoirs are the rapid formation of Jupiter[81] or, alternatively, a pressure-maximum associated with ring structures in the protoplanetary disk[82]. However, several lines of evidence indicate that this original interpretation is no longer tenable. First, the iron isotope composition of Earth's mantle is identical to CI chondrites[13], which define a CC endmember for several isotope systems (Fig. 3). As the Earth is interpreted as an NC body[10], these data are in tension with spatial isolation of NC and CC reservoirs, and require that Earth accreted considerable amounts of CC material. Second, the apparent dichotomy between $^{54}$Cr and $^{50}$Ti, between NC and CC, can be accounted for by mixing of CAI-like refractory material to CC chondrites (Fig. 3a). Third, the cosmochemistry of silicon — a new nucleosynthetic tracer — shows that all NC achondrites and Mars record identical deficits in $\mu^{30}$Si of approximately –10 ppm relative to Earth, whereas NC and CC chondrites have excesses $\mu^{30}$Si values ranging from $7.4 \pm 4.3$ ppm to $32.8 \pm 2.0$ ppm[83] (see Box 2 for a definition of $\mu$). Thus, the major division in Si isotopes is found between the parent bodies of NC-differentiated meteorites and that of NC and CC chondrite meteorites, indicating the existence of an achondrite–chondrite dichotomy as opposed to an NC and CC dichotomy (Fig. 4).

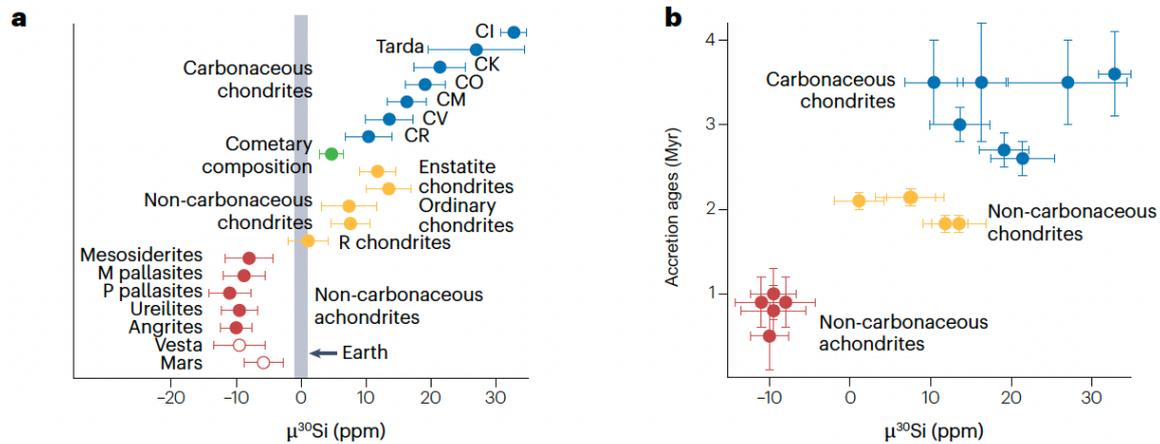

**Fig. 4 | The cosmochemistry of silicon. A.** Bulk meteorite $\mu^{30}$Si compositions (where μ is the deviations in isotopic compositions of elements relative Earth in parts per million). The vertical grey bar represents the terrestrial value derived from terrestrial basalt standards. **B**, $\mu^{30}$Si compositions versus their accretion ages. Accretion ages in Myr after calcium-aluminium-rich inclusion (CAI) condensation and are taken from ref.[227] Note that 'M pallasite' represents main group pallasite and 'P pallasite' is for pyroxene pallasite. CI: Ivuna-type chondrites; CK: Karoonda-type chondrites; CO: Ornans-type chondrites; CV: Vigarano-type chondrites; CM: Mighei-type chondrites; CR: Renazzo-type chondrites; R: Rumuruti-type chondrites. Data from refs.[28,83].

Critically, the $\mu^{30}$Si values of asteroidal bodies correlate with their accretion ages, indicating a progressive change in the composition of the disk material. These data are thus supportive of models proposing that the variability observed in the NC bodies reflects progressive admixing of outer, CI-like dust to the inner disk reservoir via inward drift of pebbles during the lifetime of the protoplanetary disk[13,21]. Admixing of outer disk material to the inner disk requires that the gaps formed by Jupiter and Saturn in the protoplanetary disk must have been leaky enough for a substantial fraction of the pebbles to pass over the low-density gap. Drift models that include a size distribution of dust aggregates indeed demonstrate that fragments formed in collisions between pebbles are advected with the gas over the gap, where they recoagulate to pebble sizes[84]. The filtering efficiency depends on a number of parameters including the fragmentation threshold speed, the strength of the turbulence, the gap depth and the gap location[85] as well as on the combined effect of multiple and possibly overlapping planetary gaps[86]. Furthermore, volatile ices may experience UV photodesorption while passing the gap[87]. Up to a third of young, massive stars show signatures of strong photospheric depletion of both volatile and refractory elements[88]; a similar depletion has been observed in the innermost regions of the evolved protoplanetary disk around the solar-mass star, TW Hya[89,90]. This provides key observational evidence that drifting dust and ice grains are trapped by pebble accretion and/or experience trapping at planetary gaps on the way towards the star. Indeed, we will later argue that a moderate reduction of the pebble-flux towards the star may have been key to forming terrestrial-planet-mass objects in the Solar System rather than super-Earths and mini-Neptunes.

Dauphas *et al.*[91] argued that the $\mu^{30}Si$ variations reported by Onyett *et al.*[83] are not nucleosynthetic in origin, but instead reflect an inappropriate correction for natural mass-dependant fractionation (expressed by the $\delta^{30}Si$ value). However, because the observed ~40 ppm range in $\mu^{30}Si$ values across asteroidal and planetary bodies is defined by objects that show no measurable $\delta^{30}Si$ variability, this potential concern only applies to the EH-type (which has a high metal content) enstatite chondrites, which are characterized by a light $\delta^{30}Si$ composition relative to most chondrites. Correcting the measured $\mu^{30}Si$ value of enstatite chondrites down to 4±3 ppm, which is still resolved from Earth, assumes that the $\delta^{30}Si$ variations between meteoritic and planetary objects reflect high-temperature vapour fractionation processes[92]. However, as pointed-out by Hin *et al.*[61], a condensation origin for the observed $\delta^{30}Si$ variability is inconsistent with magnesium isotope compositions of chondrites and Earth. Thus, we interpret the $\mu^{30}Si$ variability as nucleosynthetic in origin, such that silicon isotopes can be used to establish that chondrite parent bodies, including enstatite chondrites, are not important planetary building blocks.

In graphical representations of NC versus CC nucleosynthetic data, the lack of intermediary compositions is often cited as evidence for the lack of mixing between these two reservoirs. However, such a distribution is, in fact, predicted by models ascribing the nucleosynthetic variability amongst NC bodies by progressive addition of CI-like dust to an NC inner disk reservoir during the protoplanetary disk lifetime. Based on the $\mu^{48}Ca$ nucleosynthetic tracer, Schiller *et al.*[21] inferred that more than half of the mass of the inner disk was already locked into sizeable bodies with an NC (ureilite-like) $\mu^{48}Ca$ value formed by streaming instability about 100,000 years after the collapse of the proto-Sun — consistent with disk models[93]. Taking mass balance into account, the mixing of pristine CI-like dust from the outer Solar System with this inner-disk reservoir can, at most, raise the bulk $\mu^{48}$Ca value of the disk to approximately the terrestrial composition, thereby preserving an isotopic contrast between NC and CC bodies. The same argument can be made for other nucleosynthetic tracers, including the apparent deficit in $\varepsilon^{95}$Mo observed in the NC relative to the CC correlation line. We conclude that spatial isolation of inner and outer disk reservoirs is not required by existing meteorite data and disk models.

***Origin of the nucleosynthetic variability***

Two contrasting classes of models have been put forward to explain the Solar System's nucleosynthetic variability. In one model the nucleosynthetic isotope contrast between NC and CC bodies is ascribed to a compositional change of envelope material accreting to the protoplanetary disk in concert with outward transport of early accreted material by viscous disk expansion[66,94]. The isotopic composition of the early accreted material is hypothesized to match that of CAIs[95]. Following disk expansion, resulting in an isotopically CAI-like outer disk, the inner disk (that is, <3 a.u.) is replenished by the late addition of material characterized by an NC composition. Thus, this model predicts a CAI-like nucleosynthetic isotope signature for the outermost disk, including the comet-forming region. The discovery of primitive material from the comet-forming region preserved in a pristine meteorite provides a means to evaluate this prediction. van Kooten *et al.*[28] analysed the Si, Cr, Fe, and Mg nucleosynthetic composition of several cometary clasts, which shows that their composition is clearly not CAI-like. This observation rules out an early compositional change in the envelope material accreted to the inner disk as a cause of the observed nucleosynthetic variability. Moreover, we note that numerical simulations of collapsing clouds do not support early heterogeneity in the composition of the envelope material accreted to the disk[96].

In the competing model, thermal processing of CI-like dust results in the unmixing of different nucleosynthetic carriers, namely a thermally labile component formed in neutron-rich stellar environments, from one that is galactically inherited and thermally robust[8]. In this view, the thermally labile component represents an addition of supernova-derived nuclides to the nascent molecular cloud. Elevated temperatures in the inner disk result in the preferential destruction and volatilization of the thermally labile component and, hence, the establishment of the inner Solar System NC composition (Fig. 5). The thermally labile component has been hypothesized to represent interstellar ices mantling silicate grains[97]. CAIs are interpreted to condense from the complementary gaseous reservoir enriched in supernova-derived nuclides by volatilization of the thermally labile component. Destruction of interstellar stardust by thermal processing has also been put forward to account for the enrichment in heavy *s*–process nuclides in the NC reservoir[75]. Stellar outbursts[97] have been proposed as an efficient mechanism to thermally process inwardly drifting icy grains and, hence, generate nucleosynthetic variability, which can account for the observed nucleosynthetic variability for most elements, including the offset between the NC and CC line in Mo isotope space. The strength of this model is that it links observable disk processes with evidence for thermal processing of disk material in meteorites to nucleosynthetic variability. This, in turn, can account for the isotopic composition of CAIs, as well as the variability amongst and between NC and CC bodies.

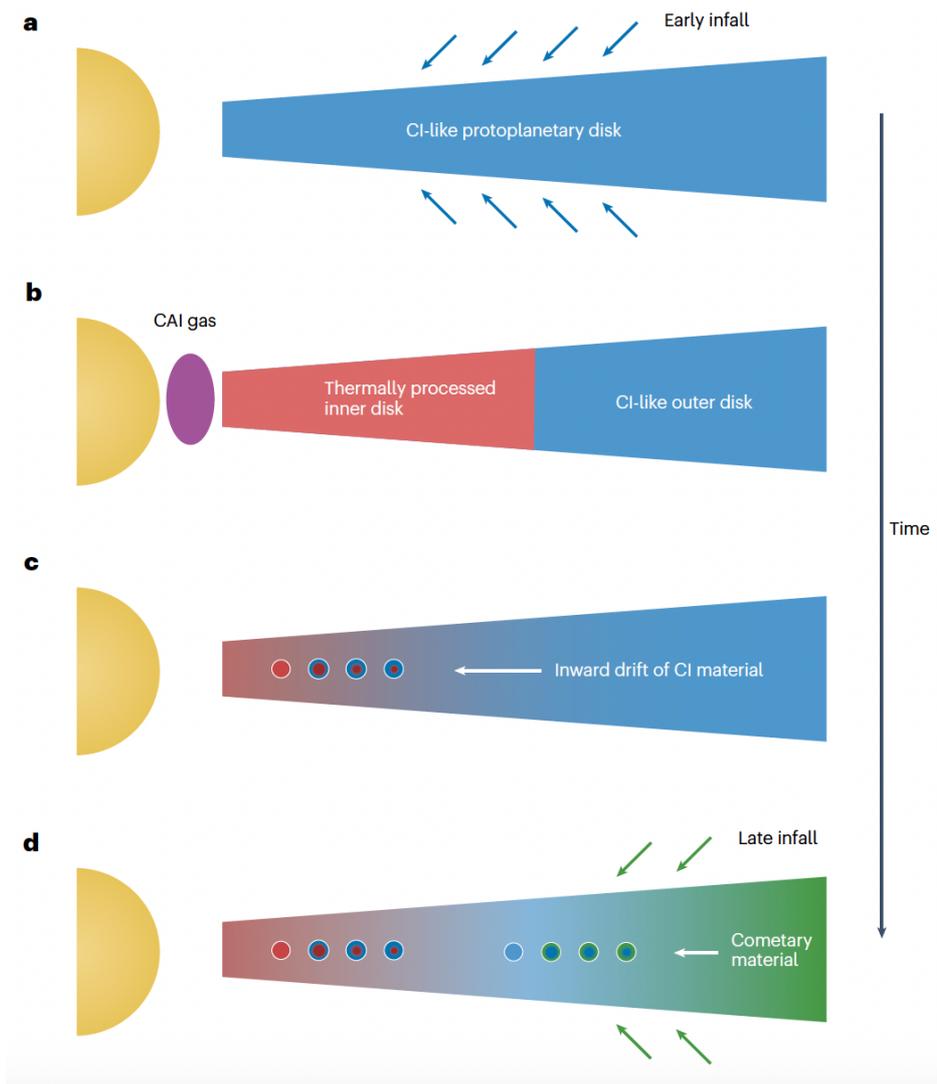

**Fig. 5 | Isotopic evolution of the protoplanetary disk. A**, Formation of the proto-Sun and solar protoplanetary disk with a CI-like isotopic composition, which represent the composition of the early infall. **B**, Thermal processing in the inner disk produces an inner NC composition and a complementary CAI gas reservoir enriched in supernova components. **C**, Inward drift of the outer CC disk material mixes the NC and CC compositions and results in NC bodies accreting variable proportion of CC material, explaining the isotopic variations amongst NC bodies. **D**, Late infall of cometary material in the outermost protoplanetary disk. Inward drift of the cometary material results in the progressive accretion of cometary material in all classes of CC chondrites (see Fig. 3F). Note that the late infall of cometary material occurs after the main accretion phase of NC bodies in the inner Solar System. CAI: calcium-aluminium-rich inclusion; NC: non-carbonaceous chondrites; CC carbonaceous chondrites; CI: Ivuna-type chondrites.

*A compositionally distinct outer disk*

Novel nucleosynthetic tracers that are insensitive to the nugget effect, associated with the incorporation of refractory inclusions (Ti and Ca) or SiC grains (Mo, Zr), can be used to define the composition of primary disk reservoirs and explore disk mass transport processes. Figure 3f shows the $\mu^{54}$Fe and $\mu^{30}$Si of NC and CC bodies together with the composition of the outermost protoplanetary disk defined by objects originating from the comet-forming region[28]. Two distinct correlations are observed, namely one between CI and inner disk NC bodies, and one between CI and CC bodies, with the cometary composition defining an isotopic endmember. As discussed earlier, the inner disk correlation is best explained by the progressive admixing of CI-like dust to a thermally processed inner Solar System reservoir[8]. In contrast, the outer disk correlation requires admixing of a distinct nucleosynthetic component represented by the cometary composition to a CI-like reservoir to account for the variability present in carbonaceous parent bodies. This suggests that, excluding CI bodies, all carbonaceous chondrite groups contained varying amounts of material from the comet-forming region, indicating that their accretion regions were not isolated.

Based on existing data, three distinct nucleosynthetic components were sampled by asteroids formed during the lifetime of the protoplanetary disk (Fig. 5), namely: the bulk Solar System composition (CI chondrites); the primordial molecular cloud composition (material from the comet-forming region); and the thermally processed inner disk (NC bodies). As CC bodies accreted late relative to NC bodies, the cometary nucleosynthetic component must represent a new addition of material to an initially CI-like disk. The primordial molecular cloud material composition is interpreted to reflect the nucleosynthetic makeup of the molecular cloud before the last addition of stellar-derived $^{26}$Al. As discussed in van Kooten *et al.*[28], this interpretation supports the supernova-trigger hypothesis for the formation of the Solar System[98]. In detail, the shockwave from a supernova event triggers the gravitational collapse and injection of gas and dust into the collapsing cloud core, parental to the Solar System, thereby accounting for the former presence of $^{26}$Al and other extinct short-lived radionuclides[99]. In this model, CI chondrites represent the bulk composition of the cloud core polluted by supernova-derived $^{26}$Al (that is, early infall). In contrast, the cometary nucleosynthetic component represents the composition of the ambient molecular cloud unpolluted by a supernova event and, hence, the primordial molecular cloud material. A possibility is that the cometary nucleosynthetic component represents material delivered to the outer disk by an accretion streamer, that is, late infall. Astronomical observations[100] and numerical simulations[101] show that streamers funnel fresh material to outer protoplanetary disks (>150 a.u.) derived from their larger-scale environments following the class 0 phase, which represent the earliest stage of stellar evolution. Thus, the nucleosynthetic make-up of the material delivered by streamers may be distinct from that of the original protoplanetary disk. Moreover, because considerable mass can be accreted via late infall, the bulk of the outer Solar System could be dominated by the cometary nucleosynthetic component (Fig. 5). This appears consistent with the observation that this isotope signal is ubiquitous in CC meteorites. Finally, the lack of the cometary nucleosynthetic component in NC bodies (Fig. 3f) may reflect the long inward drift timescales for material initially located at large orbital distances. Indeed, about 5 Myr is needed for 100 μm-sized particles initially located at 100 a.u. to reach the inner disk[102], which is comparable to the protoplanetary disk lifetime[103].

*Nucleosynthetic make-up of terrestrial planets*

When considering neutron-rich isotopes of iron group elements such as $^{48}$Ca, $^{50}$Ti, $^{54}$Cr, and $^{70}$Zn, the terrestrial composition typically lies between the NC and CC groups, suggesting that the Earth represents a mixture of NC and CC compositions. This observation has led to the development of models where the isotopic composition of the protoplanetary disk material evolves with time, such that the final composition of a body is a function of its accretion age[13,21,83]. In this view, the initial composition of the disk material is inferred to be akin to ureilite meteorites and progressive admixing of CI-like material to the inner disk results in a progressive evolution of the composition of protoplanetary disk and, hence, that of NC bodies. This model has been challenged based on the observation that Earth typically defines a compositional endmember for heavier nucleosynthetic tracers such as $^{96}$Zr, $^{94}$Mo, $^{100}$Ru, and $^{146}$Nd (ref.[20]). These data have been used to infer that Earth and Mars did not accrete inward drifting material form the CC reservoir and, as such, that the terrestrial planets did not form by pebble accretion. However, the isotopic composition of these trace elements in planetary material are highly sensitive to the variable incorporation of refractory SiC grains, which are enriched in *s*–process isotopes. In a new analysis of the nucleosynthetic composition of the most refractory planet-building element, namely silicon, Onyett *et al.*[83] showed that the terrestrial $\mu^{30}$Si composition is consistent with Earth and Mars having accreted about 30% and 10% of CI-like material, respectively, in good agreement with inferences from neutron-rich isotopes of iron group elements[21]. Importantly, the $\mu^{30}$Si values of Earth and Mars exclude NC chondrite asteroids as planetary precursors, and rather suggest that differentiated asteroids from the NC reservoir represent an important planetary building block (chondrites also yield a poor match to the terrestrial abundance of volatile elements, see Fig. 2). Collectively, the nucleosynthetic composition of Earth and Mars requires accretion of a high fraction of CI material (Fig. 6), consistent with their formation by pebble accretion during the protoplanetary disk lifetime.

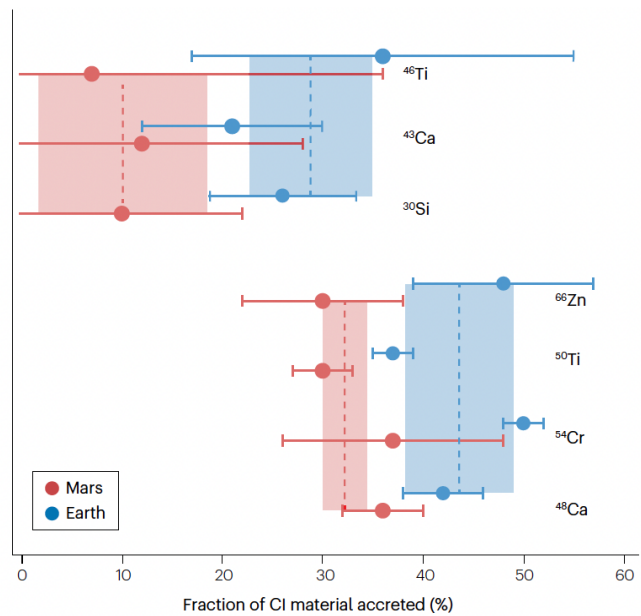

**Fig. 6 | Relative amount of CI-like material required to mix with an NC composition represented by ureilites for different isotope systems.** Note that these tracers are inferred to track the entire accretion history of a planet. The amount of CI material needed to explain the composition of Earth and Mars is higher for neutron-rich iron group elements relative to $^{46}$Ti, $^{43}$Ca and $^{30}$Si, likely reflecting distinct carriers for these nuclides. These data demonstrate that significant amounts of CI material (>20%) are needed to explain the terrestrial composition. NC: non-carbonaceous chondrites; CI: Ivuna-type chondrites.

# Making terrestrial planets

In concert with meteorite data, astronomical observations of young stars yield crucial information on the formation of planets around other stars. Observations of these stars at mm and cm wavelengths (with for example the Atacama Large Millimeter Array (ALMA) and Very Large Array (VLA) telescopes) reveal the ubiquitous presence of dusty protoplanetary disks[4]. The mass of solids in protoplanetary disks is inferred from the total stellar emission and reveals a steady decline from a few hundred $M_\oplus$ (where $\oplus$ refers to Earth) shortly after formation of the host star[104,105] to about 0.1–10 $M_\oplus$ in a few Myr[106]. Importantly, the dependence of the emission on the wavelength of the radiation gives clues to the particle sizes responsible for the emission in these disks. Direct match of the opacity index using both mm and cm data yields maximum sizes around 1 mm in the 10–100 a.u. region[107]. Thus, the making of terrestrial planets requires forming fully grown rocky bodies of several thousands of kilometres in diameter starting from mm-sized solids (Box 3).

The growth from mm-sized pebbles to 100-km diameter planetesimals is hampered by the poor sticking properties of pebbles. Laboratory experiments find a sticking threshold for silicate dust aggregates at colliding speeds of approximately 1 m s$^{-1}$ (ref.[108]), which yields maximum sizes of mm–cm for realistic protoplanetary disk conditions[109,110]. The sticking threshold for water ice at low temperatures is likely in a similar range, based on surface energy measurements[111,112]. Another issue is that collisional growth of mm-sized solids to planetesimals is too slow to compete with the rapid depletion of the protoplanetary disk solids by radial drift[102,113]. This problem, however, can be overcome by the streaming instability, which leads to the spontaneous formation of dense particle filaments in its non-linear evolution[16,114]. The filaments subsequently collapse to form planetesimals with a characteristic size of approximately 100 km (refs.[18,115–117]). How the streaming instability operates in detail in protoplanetary disks is still under exploration, particularly the role of the particle size[44,118], the particle size distribution[119], and the random, chaotic motions of gas (that is, background turbulence)[120–123]. However, the characteristic size distribution of asteroids[18,124], the large binary fraction and binary orbits of Kuiper belt objects[125], as well as the shape of the two lobes of the contact binary Arrokoth visited by the New Horizons mission[126,127] provide compelling agreement with simulations of planetesimal formation via the streaming instability.

The planetesimal model for terrestrial planet formation took inspiration from the presence of the remnant planetesimals in the asteroid belt between Mars and Jupiter[128]. Such a population of planetesimals will undergo mutual collisions and grow gradually to protoplanets[129–131]. In this view, giant impacts between these protoplanets led to the formation of terrestrial planets over a time-scale on the order of 100 million years[132], and volatiles were delivered to Earth later from icy or hydrated asteroids originating in the outer asteroid belt[133]. The lack of iron and extensive volatile depletion in Earth's moon provide compelling evidence for the giant impact hypothesis and, thus, that giant impacts must have occurred during the assembly of the terrestrial planets[134,135]. The short-lived $^{182}$Hf-$^{182}$W decay system can be used to estimate the core formation age of a differentiated body (Box 1). A high excess of $^{182}$W relative to $^{184}$W in a planetary body implies early differentiation, and vice versa if the excess is low. The low $^{182}$W excess of Earth is consistent with a moon-forming giant impact that occurred after ~50 million years[136], which is often cited as an argument to support a slow collisional growth of Earth. A giant impact has also been used to explain the high core fraction of Mercury[137]. However the lack of depletion of moderately volatile elements is at odds with the giant impact scenario[138], and the composition of Mercury may thus instead be due to its assembly from iron-rich planetesimals formed close to the Sun[139,140].

    Evaluation of the nucleosynthetic composition of Earth and Mars requires that these bodies accreted a considerable fraction of outer Solar System material (Fig. 6), which at face value is not consistent with planetesimal-driven terrestrial planet formation. The alternative view is that the terrestrial planets grow by the rapid accumulation of pebbles during the protoplanetary disk lifetime. In this model, early accreted inner disk asteroids and protoplanets acquire an NC nucleosynthetic fingerprint and progressively evolve towards the CC composition by continued accretion of inward-drifting pebbles from the outer Solar System — as such, their final composition is an intermediate between the NC and CC nucleosynthetic endmembers. Note that in pebble accretion simulations[19,141], the refractory element budget of the pebbles is not predicted to be modified by radial drift given the low midplane disk temperatures. As discussed above, several hundred $M_\oplus$ of pebbles drift inwards during the protoplanetary disk lifetime. Lambrechts *et al.*[142] demonstrated that the pebble flux through the inner protoplanetary disk determines whether planetary growth by pebble accretion is limited to protoplanet masses or whether it continues to build systems of super-Earths and mini-Neptunes, that migrate to form resonant chains anchored at the inner edge of the protoplanetary disk. The observed spacing of super-Earths and mini-Neptunes in resonant systems agrees well with standard Type I disk migration[143], which refers to the process where low- to intermediate-mass planets (typically Earth- to Neptune-sized) move through a protoplanetary disk due to gravitational interactions with the surrounding gas. The majority of observed exoplanetary

systems are not in low-order resonances[144], but this may be due to the inherent instability of resonant chains after the dissipation of the protoplanetary disk[145]. Johansen *et al.*[19] further demonstrated that the masses and orbits of the rocky planets Venus, Earth, and Mars are consistent with growth dominated by pebble accretion, starting from a primordial planetesimal belt around the current location of Mars. In this view, Mars is small because its pebble accretion growth was stunted early, while proto-Earth and Venus continued to grow and migrate to their current orbits. As already mentioned, the proto-Earth later endured a giant impact with an additional rocky planet located between the proto-Earth and Mars – dubbed Theia – that formed the Earth-Moon system. The absence of a large $^{182}$W excess in Earth's mantle predicted by the rapid formation and differentiation of the proto-Earth during the disk lifetime appears to be an argument against pebble accretion. However, partial equilibration of Theia's core with the mantle of the proto-Earth during the giant impact can reset the original W isotope composition to its observed value[146,147], even for a relatively low equilibration level of ~25%. This reset to the terrestrial value nevertheless requires a nearly-equal mass impactor, which deviates from the canonical model of a Mars-sized impactor[148]. Although the size of the Moon-forming impactor is debated[149], we emphasize that the Hf-W system is also consistent with Earth accreting 70% of its mass by pebble accretion, followed by an extensive bombardment with large impactors for the next 40 Myr and finally experiencing an impact with a Mars-sized protoplanet at a more conservative 10% equilibration efficiency[147]. Thus, pebble accretion may be an important driver of planetary growth in the inner regions of the protoplanetary disk, and planetary impacts clearly also play a role in shaping systems of terrestrial planets, super-Earths, and mini-Neptunes.

Mercury, Venus, Earth, and Mars have all differentiated from other planets to form a metal core and a mantle dominated by silicate rock. This implies that rocky planets melt during their accretion[150]. Consecutive giant impacts can melt a large fraction of the mantle, causing metal to sink to the core[151,152]. In contrast, the accretion of pebbles leads to direct differentiation through trapping of the accretion energy in an early-accreted water atmosphere. This results in run-away greenhouse heating that melts the growing planet from the top down after the planet reaches a mass of 0.01 $M_\oplus$[146,153]. Volatiles such as $H_2O$, C, and N are accreted as parts of both dry and hydrated planetesimals[133], and pebbles with rims of water-ice and organic molecules[19,154]. Thus, in pebble accretion, volatiles are acquired early on, during the main accretion phase of the planet as opposed to later on by impacts from colliding water-rich asteroids. These volatiles dissolve in both silicate melts and metal melts, with a partition coefficient that depends on temperature and pressure[155]. H and C are siderophile (having an affinity for iron) and partition more strongly into metal than into silicates[156,157]. Thus, the planetary core likely contains the majority of the accreted H and C budgets[154]. N is also strongly siderophile[158,159], but it dissolves poorly in magma to begin with[160] and its abundance in the core could therefore be very low since N will mainly reside in the atmosphere[154,161].

**Thermal processing of planetary building blocks during pebble accretion**

Protoplanets that grow to masses larger than approximately 0.01 $M_\oplus$ within the protoplanetary disk attract a hydrostatic envelope consisting primarily of $H_2$ and He (refs.[19,162,163]). The temperature rises dramatically in this envelope closer to the surface of the protoplanet. Thus, pebbles that traverse the envelope are exposed to thermal processing and loss of volatiles. Wang *et al.*[164] demonstrated using hydrodynamical simulations how ultravolatiles such as $H_2O$, which sublimate relatively far out in the Hill sphere, are returned to the protoplanetary disk via gas flows that penetrate from the disk through the Hill radius and back (so-called recycling flows[165]). Johansen *et al.*[19] used such a mass loss mechanism to propose that bulk Earth may have accreted a considerable fraction of its mass by pebble accretion exterior to the water ice line and that more than 95% of the pebbles' ice was lost during accretion. Moderately volatile elements, on the other hand, sublimate within the deeper, convective parts of the envelope. Refractory organic molecules are processed in these layers to release mainly the ultravolatile molecules $NH_3$ and $CH_4$ as decomposition gases (refs.[166,167]), while FeS reacts with $H_2$ to release volatile $H_2S$ (ref.[168]). These species escape swiftly by turbulent diffusion into the outer regions of the Hill sphere where they enter the recycling flows[19,169]. Thermal processing of FeS to form escaping $H_2S$ gives a good match to the strong bulk sulfur depletion of Earth and moderate depletion of Mars[169]. We note that in this model, the sulfur addition from the moon-forming giant impact is key to avoiding overdepletion of Earth's sulfur content. Elements more refractory than Si are prevented from escaping to the recycling zone by the formation of a dense layer of SiO-dominated vapor in equilibrium with the underlying magma ocean, which is separated from the overlying convective envelope by a radiative zone[170]. This may explain why Earth could retain a CI-like level of the elements more refractory than Si (Fig. 2). Alternatively, in the planetesimal-driven model for terrestrial planet formation, the depletion of moderately volatile elements such as S could be a consequence of planetesimal formation in an early, hot stage of the solar protoplanetary disk[171].

In addition to volatile depletion, thermal processing of pebbles in the hot gaseous envelope surrounding an accreting protoplanet can further impart nucleosynthetic variability by selective destruction of volatile carriers, in a similar fashion as proposed for thermal processing of dust in the inner protoplanetary disk[8,75]. Earth can be plotted as an *s*–process enriched compositional endmember amongst NC bodies for several nucleosynthetic tracers such as Mo, Nd, and Zr, an observation used to argue that Earth formed from material unsampled in the meteorite record[20]. However, because these nucleosynthetic tracers are highly sensitive to the variable incorporation of *s*–process enriched refractory SiC grains, thermal processing of accreting pebbles in the hot planetary envelope must be considered. Onyett *et al.*[83] explored the effect of thermal processing of CI-like material in the planetary envelope during Earth's formation by pebble accretion. In these simulations, the Solar System's *s*–process variability reflects unmixing of two dust reservoirs, namely, homogenized interstellar dust and an *s*–process-dominated SiC stardust component[75]. Because sulphides (FeS, MgS, and CaS) are an important reservoir for Mo, Nd, and Zr, sublimation of sulphides in the planetary envelope starting at ~690 K will result in the losses of these trace elements, enriching the pebble residues in refractory SiC grains. Onyett *et al.*[83] calculate that a modest 10% loss of the Mo, Nd, and Zr budget from the interstellar dust reservoir during thermal processing in the envelope can reproduce the terrestrial composition. Thus, Earth's *s*–process enrichment relative to achondrite and chondrite parent bodies could be a hallmark of its different accretion history, namely, pebble accretion as opposed to streaming instability, and collisional growth.

The cosmochemistry of the moderately volatile element Zn has been developed to probe the nature of the building blocks or Earth and Mars[69–71,172,173]. Based on the relationship between the $\mu^{54}$Cr and $\mu^{66}$Zn tracers, Kleine *et al.*[173] suggested that both Earth and Mars must have accreted only minor amounts of CC material (<10%), an interpretation that is inconsistent with constraints from other refractory nucleosynthetic tracers such as Cr, Ti, Ca, and Si (Fig. 6). However, because Zn is much more volatile than Cr (~50% condensation temperatures of 723 K and 1,291 K, respectively[174]), Zn loss by volatility-driven processes needs to be carefully evaluated. Volatile loss has been simulated experimentally by progressively heating CM-type chondrites (Mighei-type carbonaceous chondrites) and tracking Zn depletion as a function of temperature[175]. These experiments clearly demonstrate that over 99% of the original Zn budget is lost before reaching ~800 K. Similarly, strong Zn depletion has also been documented in CM chondrites that have been naturally heated to similar temperatures by impact processes[176], in agreement with experimental data. Thus, substantial Zn loss is predicted to occur during envelope processing of pebbles such that the Zn and Cr content of the accreted material will be decoupled. Based on the models developed in Johansen *et al.*[19] and Onyett *et al.*[83], we can show that the $\mu^{54}$Cr-$\mu^{66}$Zn evolution of a planet with an initial NC (ureilite-like) composition grows by accretion of CI-like pebbles (Fig. 7). This highlights that Zn is only accreted during the early stage of pebble accretion before the Zn sublimination temperature is reached in the envelope. We conclude that the Zn isotope composition of Earth and Mars are consistent with high fractions of accreted CI material, in agreement with what has been inferred for refractory tracers.

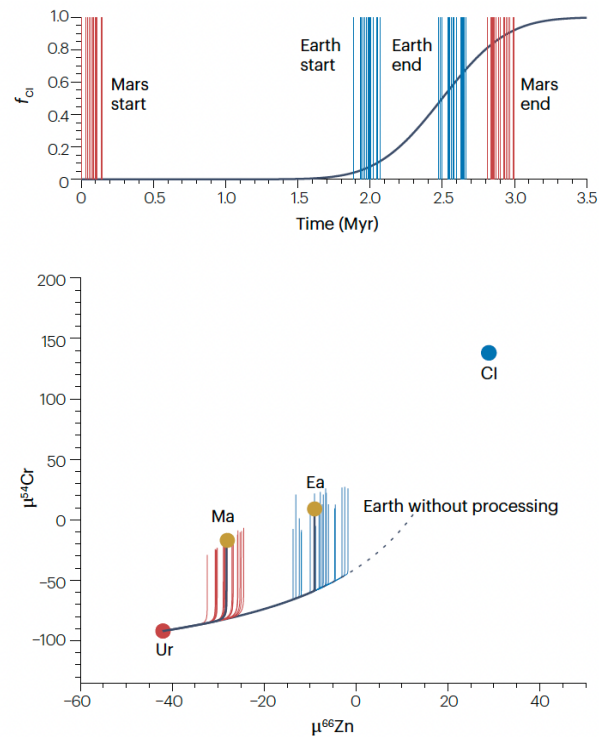

**Fig. 7 | Thermal processing and the $^{54}$Cr-$^{66}$Zn composition of Earth and Mars.** The top panel shows the evolution of the Ivuna-type chondrite (CI) fraction in the terrestrial planet formation zone, based on a stellar outburst model[97,141]. Overplotted are a range of start times and end times for accretion of Earth and Mars that give the best matches to the evolution of $\mu^{66}$Zn and $\mu^{54}$Cr as shown in the bottom panel (where μ is the deviations in isotopic compositions of elements relative Earth in parts per million). Used here, is a Zn abundance of 300 ppm for CI chondrites and 50 ppm for the inner Solar System end-member ureilites (Ur). The volatility of Zn implies that Zn can only be accreted while the protoplanet has a mass below 0.04 $M_\oplus$. Above that mass, the Zn composition is frozen while the Cr composition continues to evolve towards the CI endmember. The dashed line shows how the composition misses Earth (and Mars) if loss of Zn by thermal processing is not taken into account. For the two isotopes plotted here, Mars is interpreted to have experienced slow but extended growth over the whole disk lifetime, while Earth experienced more rapid growth during the time when the inner regions of the protoplanerary disk were enriched with CI material.

# Planetary atmospheres and potential habitability

Planetary atmospheres play a crucial role in determining the habitability of a world. Their composition, evolution, and interaction with stellar radiation influence surface conditions, climate stability, and the potential for life. The formation of planetary atmospheres is a complex process, shaped by various mechanisms such as outgassing, volatile delivery, and atmospheric loss. In the following sections, we explore the origins and characteristics of planetary atmospheres, with a particular focus on their implications for habitability.

*Atmospheric outgassing*

Planetary atmospheres can be of different origins[177]. They can be accreted from the gaseous disk, outgassed during the magma ocean crystallization, delivered by collisions with icy asteroids and comets, or outgassed from volcanoes during the long-term evolution of the solid interior. These types of atmospheres are generally distinct in their formation timescale, composition, and pressure. We highlight in this review magma ocean atmospheres as a direct legacy of the planetary formation stage, since most of the accreted volatiles are expelled during magma ocean crystallization. Magma oceans arise naturally during the pebble accretion phase when the accretion energy is trapped below the proto-atmosphere[146], as well as after giant impacts in the post-disk phase such as the moon-forming impact[178]. The presence of a terrestrial magma ocean already during the protoplanetary disk phase is supported by the presence of Ne with nebular isotopic composition in the deep mantle[179]. The magma ocean of Mars, on the other hand, must have crystallized within 20 Myr of planet formation based on the timing of extraction of the primordial crust[48].

Traditionally, magma ocean outgassing models have focused on the physics of a cooling planet. Only since the 2020s have chemical reactions within the magma ocean received the necessary attention in models for comparative planetology. Critically, the redox state of the magma ocean greatly influences the chemistry of the outgassing species[180]. The redox conditions of magma oceans control the redox state of subsequently formed mantles and the overlying atmospheres[181]. Oxidised magmas outgas species like $H_2O$, $CO_2$, and $SO_2$, whereas reduced magmas release $H_2$, CO, $H_2S$, and small amounts of $H_2O$ or $CH_4$ (ref.[182]). As a magma ocean solidifies, its melt fraction decreases, and most volatiles are outgassed. The outgassed volatile speciation evolves during cooling. Volatiles with low solubility (for example, CO and $CO_2$) are outgassed to the surface first, leading to a temporary carbon-rich early atmosphere during cooling, followed by the outgassing of the more soluble species like $H_2O$[183].

Earth's magma ocean must have been highly reducing during accretion when it was still in equilibrium with free metals in the silicates and in the core[184]. However, investigations of Earth's earliest minerals indicates that, between 4 and 4.4 billion years ago, the upper mantle was already highly oxidised, similar to the modern mantle[185]. This oxidation is now attributed to the disproportionation of ferrous iron ($Fe^{2+}$) to ferric iron ($Fe^{3+}$) and metallic iron ($Fe^0$) in the magma ocean stage, after metallic iron had sequestered into the core but before the final crystallization. Specifically, the redox buffering reaction $FeO(melt) + \frac{1}{4} O_2 = FeO_{1.5}$ (melt) is responsible for oxidizing ferrous iron to ferric iron. As this reaction is pressure dependent, deeper magma oceans have considerably higher ferric iron contents, explaining the highly oxidized Hadean mantle[186].

Although the relevant redox buffering reactions are still being investigated, thermodynamic models indicate a clear trend of redox state with planet mass caused by the pressure-dependence on redox reactions[186]. Thus, early atmospheres in equilibrium with magma oceans are dominated by $CO_2$ and $H_2O$ for giant planets, whereas smaller bodies are predicted to outgas more reduced gases early in their evolution, like $H_2O$, $H_2$, and CO for Mars, and mainly $H_2$ and CO for the Moon. Outgassing of large amounts of $H_2$ makes small planets like Mars susceptible to complete loss of the primordial atmosphere by extreme ultraviolet (XUV) irradiation[154,187] as the escaping H atoms drag along with them heavier species such as C, N, and O.

The iron-dominated cores of planets constitute an important reservoir of volatiles. During planet formation, large-scale melting allows the iron to separate from the silicates to form a core. Iron-melt droplets that sink through the silicate-dominated magma ocean equilibrate with the silicates — and this allows volatile species to also equilibrate between iron and silicates. For Earth-like magma conditions, hydrogen and water partition predominantly into the iron[156,188]. This partitioning is even stronger for higher-mass planets[189], suggesting that water-rich worlds naturally incorporate water in their cores. This has important consequences for the volatile budgets of rocky planets, as species sequestered in the core likely remain bound there for the entirety of planetary evolution[154]. Refractory siderophile species also partition into metal melts. Because the silicate-metal partition coefficients are dependent on temperature and pressure, it is possible to infer the mean equilibration pressure from the measured abundances of siderophile species[155]. The inferred pressure agrees broadly with the deep magma oceans expected from both pebble accretion and giant impacts[153]. However, note that planets generally start their long-term evolution under hot conditions with deep global magma oceans.

For close-in exoplanets, their atmospheres can undergo further oxidation via preferential loss of hydrogen to space. For example, water vapour at high altitudes is subject to photolysis followed by loss of hydrogen to space and enrichment in oxygen[190]. When loss rates are high, fractionation of gas species are not expected as the hydrodynamic outflow drags away the bulk of gas. Only when loss rates are low can hydrogen be preferentially lost compared to other species. In summary, volatiles are accreted together within the planetary building blocks and partition between the core, the mantle, and the atmosphere — a considerable part of the atmosphere may subsequently be blown off by stellar XUV irradiation.

*Atmospheres of small exoplanets*

The majority of observed exoplanets have radii ($R$) less than 3.5 $R_\oplus$ and orbital periods less than 100 days[191]. Understanding how these planets formed has profound implications for generalizing models of rocky planet formation to exoplanetary systems. At around 1.9–2 $R_\oplus$, there is a clear scarcity of planets[192,193]. Sub-Neptune classes (1.9–3.5 $R_\oplus$) have thick H and He atmospheres, while super-Earths (1–2 $R_\oplus$) either lost their atmospheres[194–196] or never accreted them in the first place[197]. These super-Earth planets have silicate-to-iron ratios similar to their host stars[198–201]. The composition of small planets is nevertheless quite degenerate[202,203] as planets of different interior compositions can have identical mass and radius.

Transit spectroscopy is a promising pathway to reveal the chemical properties of rocky planet atmospheres. In the new era of using the James Webb Space Telescope (JWST), there is a growing number of high-precision spectroscopy targets becoming available. Among the most temperate and small planets is the Trappist-1 system, where Greene *et al.*[204] ruled out a considerable atmosphere for Trappist-1b, and Lincowski *et al.*[205] identified a potential atmosphere on Trappist-1c. The atmospheric characterization of Earth-mass planets remains very challenging, even with modern space-based telescopes.

*Atmospheric composition and planetary habitability*

The developments reviewed in this article on planetary growth processes, volatile delivery during the accretion phase, and on the evolution of the oxidation state of magma oceans allow us to present a population synthesis map of planetary orbits, masses, and compositions (Fig. 8). We create the synthetic planet population with the mass and orbit evolution of planetary embryos from a starting mass $M = 10^{-4}$ $M_\oplus$, with randomly sampled starting locations between 0.1–10 a.u. and random starting times between 0.1–3 Myr during the evolution of the protoplanetary disk[19,206]. Importantly, we monitor the abundance of water and carbon in the synthetic planets, both in the protoplanetary disk and during the accretion. Starting from the exterior, at ambient disk temperatures below 120 K, ice sublimation occurs within the convective envelope of the growing planet. This accreted water remains bound to the planet, in analogy to how silicate vapour is protected from escape by the emergence of a radiative zone above the vapour region. These bodies grow to water-rich planets with more than 10% water mass fraction. Planets that reach the pebble isolation mass — above which pebbles are prevented from accreting onto the planet due to the emergence of a shallow gap in the protoplanetary disk[207] — attract a gas envelope, which increases in mass only very slowly through cooling and contraction[14,208]. The gas accretion rate decreases closer to the star, where the pebble isolation mass is lower. These worlds of magma, steam, and gas[209] migrate towards the inner edge of the disk. Planets that form in the temperature range between 120–170 K, on the other hand, accrete water ice only until the protoplanet reaches a mass of ~0.01 $M_\oplus$, after which water sublimation from the icy pebbles leads to efficient vapour loss back to the protoplanetary disk[19,164]. The carbon in organic species is accreted up to ~0.04 $M_\oplus$, with only a small fraction of refractory carbon surviving up to ~1,100 K. We assume that volatile carbon-bearing molecules released by thermal decomposition of organic compounds are protected from mass loss under cold conditions (<120 K) by the radiative zone that also protects the water. We mark the transition from reducing to oxidizing atmospheres in the habitable zone in the lower panel of Fig. 8 (based on ref.[210] assuming Earth-like FeO fraction in the mantle). Planets with masses above ~0.2 $M_\oplus$ undergo self-oxidation in the magma ocean[211] to outgas an atmosphere dominated by $H_2O$ and $CO_2$, while smaller planets in the same region outgas an $H_2$-rich atmosphere that is easily lost to XUV irradiation[154]. Rocky planets that migrate over the inner edge of the habitable zone undergo a run-away greenhouse effect whereby water is destroyed by XUV. Conversely, planets starting interior of the water ice line do not accrete any water and outgas atmospheres devoid of $H_2O$, but possibly rich in C and N from the accretion of refractory organic compounds that survive up to ambient temperatures of ~400 K[19,167]. Earth may thus have avoided becoming either a dry desert planet or a completely water-dominated world through its fortuitous formation location just exterior of the water-ice line in the protoplanetary disk, where thermal processing during the accretion had a major effect on reducing the volatile budget. Fig. 8 also gives a perspective on the possibility of volatile delivery via stochastic impacts of volatile-rich planetesimals or protoplanets. Small bodies forming in the 1 a.u. region as well as larger bodies that grow exterior of 1.5 a.u. are very water-rich and could deliver additional volatiles via impacts to an Earth-like planet that grows in the habitable zone.

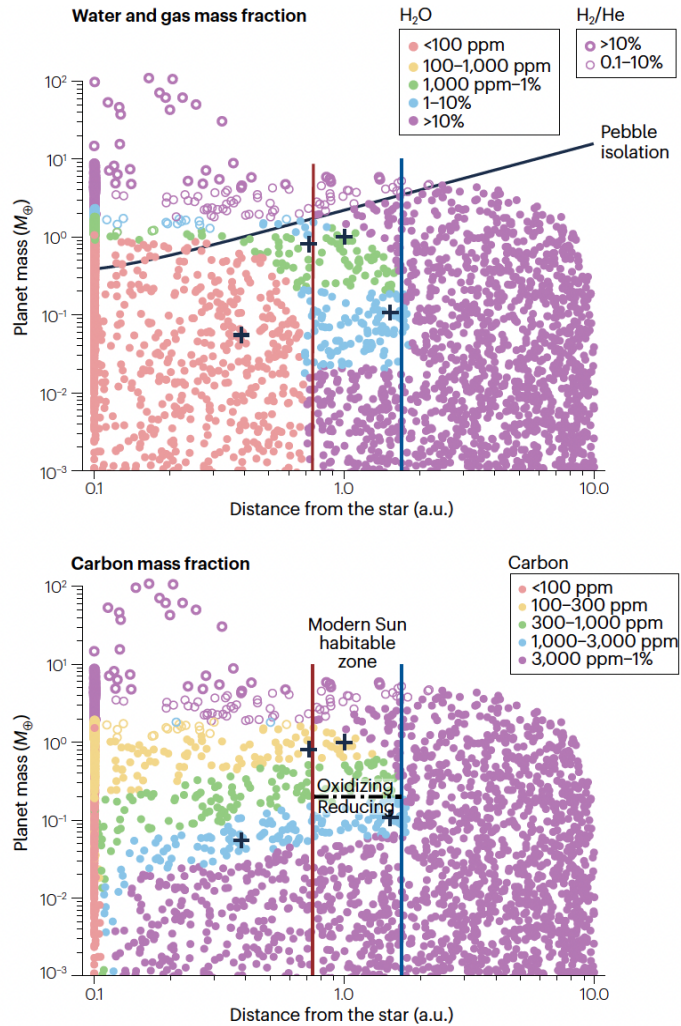

**Fig. 8 | Population synthesis calculation of planet formation in the inner regions of the protoplanetary disk.** Planetary embryos initialize between 0.1 and 10 AU and grow from an initial mass of $10^{-4}$ $M_\oplus$ by pebble accretion in a protoplanetary disk with a weak turbulent diffusion of $\alpha = 10^{-4}$ and a pebble fragmentation threshold of $v_f = 1$ m s$^{-1}$. The top panel shows the water content with color; planets that reach pebble isolation mass and accrete an envelope of $H_2$ and He are indicated with open circles. Sublimation of water vapor from pebbles during accretion manifests as decreasing water content with increasing planetary mass exterior to the water-ice line (0.7 a.u,). The bottom panel shows the carbon content. We assume that 90% of the carbon is carried in refractory organic compounds destroyed progressively between 325 and 425 K and 10% in graphite destroyed[166,167] at 1,100 K. Both water and carbon are assumed to be protected from recycling back to the protoplanetary disk under very cold conditions (<120 K), where water sublimation occurs deep in the envelope and creates a radiative zone. We set, for simplicity, the carbon content of the pebbles to 1%. For reference, the Earth's surface oceans

add up to 230 ppm and Venus' atmospheric carbon to 25 ppm relative to the mass of the planet. The bulk H and C contents of these planets must nevertheless reside in their cores, due to their siderophile nature in the magma ocean[156,157]. We indicate in the lower panel the empirical habitable zone of the modern Sun[228], as well as the transition from a reducing outgassed atmosphere (dominated by $H_2$, $H_2O$, $CH_4$ and $NH_3$) to an oxidizing outgassed atmosphere (dominated by $H_2O$, $CO_2$ and $N_2$); increased oxidation is due to self-oxidation of the magma ocean at high base pressures[210,211].

**Challenges and outlook**

The formation timescales and mechanisms of the Solar System's terrestrial planets remain highly debated in the community. Slow collisional growth in the framework of the planetesimal-driven model as well as rapid planetary growth by pebble accretion have both been put forward to explain the planetary architecture of the Solar System. Distinguishing between these models is critical because they make contrasting predictions with respect to the volatile budget of rocky planets and, hence, their potential habitability. In the standard model, the delivery of volatile elements and complex organic chemistry critical to life occurs by stochastic processes late in the history of the terrestrial planets, perhaps as late as >100 Myr after the formation of the Solar System. In the pebble accretion model, volatiles and organic species are delivered to rocky planets during their main accretion phase within the lifetime of the protoplanetary disk (a few Myr). Thus, if the growth of rocky worlds is dominated by pebble accretion, the presence of volatiles necessary for habitability is a generic feature of this planet formation mechanism.

A clear consequence of pebble accretion is that the planetary building blocks are exposed to strong thermal processing during the accretion. This thermal processing has the potential to explain the abundances of both the volatile (atmosphere-building) and the moderately volatile elements of Earth and Mars and gives clear predictions for the composition and potential habitability of rocky planets around other stars. However, the first hydrodynamical simulations that capture the fate of volatiles during pebble accretion are only now becoming possible. More advanced computer simulations will be needed to fully understand the fate of the sublimated volatiles after they are released in either the recycling zone or in the gravitationally bound regions of the planetary envelope.

An important research avenue to distinguish between planet formation mechanisms is to better quantify the extent of mass transport during the protoplanetary disk lifetime, specifically the flux of outer disk material to the accretion region of terrestrial planets. Indeed, in the interpretation that the NC-CC dichotomy reflects the early spatial isolation of inner and outer disk reservoirs, no mixing of the two compositions is permitted. In contrast, if the dichotomy represents a compositional gradient as argued in this Review, evidence for mixing should exist at the scale of mm-sized solids as well as early-accreted protoplanetary bodies. In concert with the cosmochemistry of silicon, a novel nucleosynthetic tracer[83], chondrules can be used to decipher the radial dust flux as they represent early-formed solids and are ubiquitous components of chondrites accreted in both the NC and CC regions. Silicon is a major component rock to such an extent that it is possible to determine its isotopic composition in minute inclusions. As mm-sized chondrules typically have igneous and dusty rims that represent distinct generations of material added during chondrule transport and remelting, the composition of their cores and mantling rims can be used to quantify the radial dust flux and mass transport across disk reservoirs. We note that a report on the silicon isotope composition of chondrules and their rims is consistent with mixing of NC and CC compositions[212], although these data must be expanded to all known classes of chondrites. Similarly, the nucleosynthetic composition of a large sample of first-generation asteroids can be used to determine whether a compositional gradient, as opposed to a dichotomy, exists between first-generation NC and CC bodies. In contrast to known iron meteorites groups, which represent samples from only ~10 parent bodies[213], most ungrouped irons may represent the only samples of their >100 individual parent bodies[214]. Thus, ungrouped iron meteorites sample more parent bodies than all the other meteorite types combined, representing a unique source of untapped information regarding the nucleosynthetic diversity of protoplanets formed in the first Myr of the protoplanetary disk evolution. Of particular interest is the cosmochemistry of Zn and Ge, which are novel tracers with similar nucleosyntheses that include nuclides formed exclusively by $s$– and $r$–process nucleosynthesis ($^{64}$Zn-$^{70}$Ge, $^{70}$Zn, and $^{76}$Ge, respectively). The observed dichotomy in Mo isotopes is only observed in $^{94}$Mo ($p$–process) versus $^{95}$Mo ($s$– and $r$–process) isotopes and has been interpreted as an $r$–process excess in the CC reservoir[10]. If correct, a similar pattern must exist when coupling $s$– and $r$–process nuclides of Zn and Ge. Thus, measuring the nucleosynthetic composition of Zn and Ge in ungrouped irons is timely.

Secondary processes during and after the main accretion phases of rocky planets also require careful examination. For example, the primary composition of some nucleosynthetic tracers (such as Mo and Zr) can be modified by volatility-driven processes during planetary growth by pebble accretion. Similarly, the nucleosynthetic composition of Earth's mantle for siderophile elements can be modified by equilibration with the core of the Moon-forming impactor. In concert with the development of novel cosmochemical tracers, these efforts can be used to define a unifying model that can explain the nucleosynthetic variability that exists are the scale of mm-solids, asteroids and planets.

Finally, observations of volatiles in the inner regions of protoplanetary disks are undergoing a revolution with the launch of the JWST. Connecting such observations of volatiles to radial drift models and to the composition of rocky planets is a promising avenue to better quantify volatile delivery in the habitable zone and how this depends on properties of the protoplanetary disk and the host star[215,216].

**Acknowledgements**

We thank three anonymous referees for their constructive comments. M.B. acknowledges funding from the European Research Council (ERC Advanced Grant Agreement 833275 — DEEPTIME) and the Villum Fonden (54476). A.J. acknowledges funding from the European Research Council (ERC Consolidator Grant Agreement 724687 — PLANETESYS), the Knut and Alice Wallenberg Foundation (Wallenberg Scholar Grant 2019.0442), the Swedish Research Council (Project Grant 2018-04867), the Danish National Research Foundation (DNRF Chair Grant DNRF159) and the Göran Gustafsson Foundation.

**Author contributions**

M.B., A.J., and C.D. contributed equally, and all authors were part of the discussion, reviewing, and editing of the manuscript.

**Competing interest**

The authors declare no competing interests.

**Additional information**

**Peer review information**

*Nature Reviews Chemistry* thanks the anonymous, reviewers for their contribution to the peer review of this work.

**Box 1 | Dating meteorites**

The radioactive decay systems used to date meteorites and their components can be subdivided into long-lived and short-lived chronometers, or timescales. Of the long-lived chronometric systems, only the dual decay of $^{238}$U and $^{235}$U to $^{206}$Pb and $^{207}$Pb, respectively, have appropriate half-lives (~4.5 and 0.7 Gyr, respectively) to resolve the ages of objects formed in the first 5 Myr of the Solar System[229] provided that their $^{235}$U/$^{238}$U ratios are measured[230]. Using this system, an accurate chronology was established for timing solid-formation, asteroidal melting, and subsequent solidification. The Solar System's first solids, calcium-aluminium-refractory inclusions (CAIs), define a U-corrected Pb-Pb age of 4,567.3±0.16 million years before present time, which is taken as a reference age for the start of the Solar System. Short-lived chronometers are based on nuclides that existed at the start of the Solar System but have since decayed away such that their former presence can only be inferred from the isotopic composition of their daughter isotopes[231]. Thus, these chronometers can only be used to define relative age differences between events, and a central assumption for these ages to be meaningful is that the parent nuclide was homogeneously distributed at the start of the Solar System.

   For example, the $^{26}$Al-$^{26}$Mg decay system, with a half-life of 700,000 years, has been widely used to date the formation timescale of CAIs, chondrules, and differentiated planetesimals. However, U-corrected Pb-Pb and Al-Mg ages are often not concordant for the same objects, an observation that has been interpreted by some to reflect heterogenous distribution of $^{26}$Al (refs. [40,219,232–234]) although other interpretations have also been proposed[235]. Short-lived systems with longer half-lives can date planetary processes such as differentiation, core formation, and crust solidification. The Hf–W system (half-life of ~9 Myr) is useful to date core formation because Hf and W have different geochemical behaviours such that Hf resides in the silicate part of the planet, whereas W partitions into the core as the planet solidifies. The W isotope composition of iron meteorites, which represent fragmented cores of differentiated asteroids, can be used to determine their core formation age from the start of the Solar System[236]. For large bodies like Earth and Mars, where we do not have samples of their cores, the W isotope composition of the silicate part of the body can, under reasonable assumptions, be used to provide a core formation model age[136]. For Earth, the relatively late core formation age >50 Myr after Solar System formation is often cited as an argument to support slow collisional growth of Earth as opposed to rapid formation by pebble accretion during the protoplanetary disk lifetime. However, the W isotope composition of the silicate part of a planet can be reset during large impacts by partial equilibration with the impactor core — as has been suggested for the giant impact that formed the Earth-Moon system[237,238].

It is also possible to estimate the accretion age of an asteroid by modelling the heat produced by the decay of $^{26}$Al in asteroids formed at different times[239]. This provides time-integrated temperature profiles that can be compared with the peak temperatures inferred for asteroids based on meteorite data. Although the results are model-dependent, these ages give useful information to understand the accretion history of differentiated and non-differentiated asteroids.

**Box 2 | Nucleosynthetic fingerprinting**

Chondrites contain a record of the different stars that contributed matter to the molecular cloud core, parental to our Sun, in the form of stardust grains[25]. These rare, tiny minerals (<1 μm in size) of graphite, nitride, oxides, and silicates, condensed around dying stars prior to the formation of our Sun[240]. Stardust grains have widely anomalous isotopic compositions relative to the bulk Solar System composition, which reflects specific nucleosynthetic processes in stars. Variable incorporation of this anomalous material when disk solids, asteroids, and planetary bodies formed can impart isotopic heterogeneity and, as such, the chondritic stardust inventory provides a framework to understand the Solar System's nucleosynthetic variability. Broadly, most pre-solar grains come from asymptotic giant branch (AGB) stars (the site of *s*–process nucleosynthesis) with some from diverse supernova environments. As such, heterogeneity in the abundance of isotope species formed in such stellar environments is expected between Solar System materials.

Anomalies of nucleosynthetic origins in meteorites are deviations in the isotopic composition of an element relative to that of Earth, typically expressed in deviations in parts per millions (μ notation) or parts per 10,000 (ε notation). Most nucleosynthetic anomalies have been identified in elements from and beyond iron group elements[241]. For example, the $^{48}$Ca/$^{44}$Ca ratio of asteroids and planets in the Solar System, expressed as the μ$^{48}$Ca, varies from –150 ppm to +300 ppm relative to Earth[21]. This means that each body has a distinct nucleosynthetic fingerprint, which can be used — like DNA — to explore genetic relationships and determine the nature of the precursor material to planets. Although some of these anomalies can be linked to a specific type of pre-solar grain, the carriers of most isotope anomalies in meteorites are essentially unknown and, hence, so are their thermal properties. It has been suggested that thermal processes can impart nucleosynthetic variability. This can happen in the precursors to asteroids and planets in the protoplanetary disk[8,75] or, alternatively, by volatility-driven processes during the accretion of a planet[83]. Thus, these potential pitfalls must be carefully evaluated when interpreting and comparing nucleosynthetic data from isotopes with distinct origins.

**Box 3 | Pebble accretion versus collisional growth in terrestrial planetary formation**

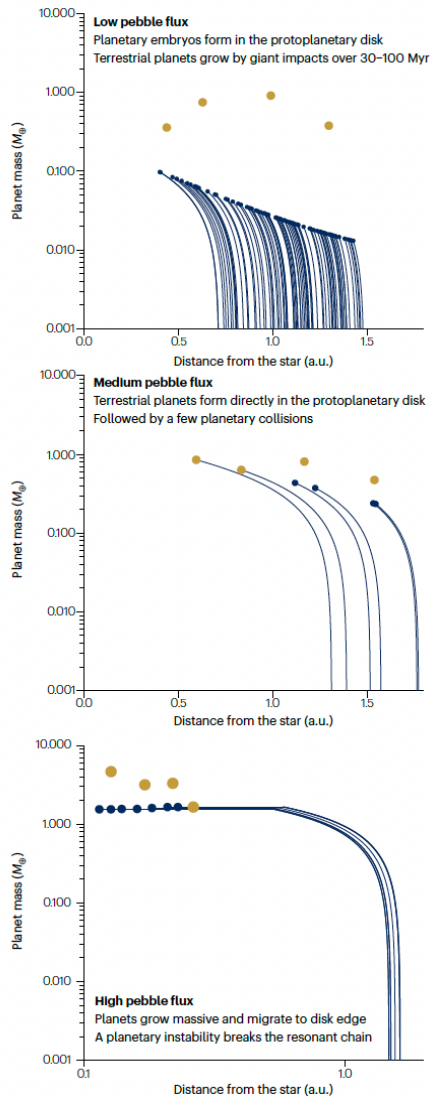

The growth of planetesimals to planetary bodies occurs through a combination of mutual impacts and accretion of pebbles. The plot above illustrates different outcomes of planet formation in the terrestrial planet zone. The maximum masses reached by pebble accretion in the inner regions of a protoplanetary disk depend strongly on the radial flux of pebbles through the zone of terrestrial planet formation[142]; a variation in the pebble flux from star to star is predicted to arise from the different filtering efficiencies caused by one or more cores and giant planets that accrete pebbles and carve gaps further out in the disk[84,85]. A low pebble flux (top panel) implies that planetary growth is stunted at the embryo stage, below 10% of the mass of the Earth, when the protoplanetary disk dissipates after a few million years. These embryos then undergo mutual impacts over timescales of 30–100 million years to form a system of terrestrial planets. A medium pebble flux (middle panel) will nevertheless allow some embryos to grow to full terrestrial planet mass within the protoplanetary disk[19]. Such a system later experiences just a few planetary collisions; the moon-forming giant impact may have been such a collision between proto-Earth and an additional rocky planet that formed between Earth and Mars. For a high pebble flux (bottom panel), planets grow within the protoplanetary disk to masses above the mass scale of the terrestrial planets and migrate towards the inner edge of the disk. Here, they pile up in resonant chains where disk migration is stopped by resonant interaction with the interior planet[142]. Most resonant chains undergo planetary instability after the dissipation of the protoplanetary disk, to form systems of hot super-Earths and mini-Neptunes[145].

**Glossary**

**Primitive asteroid**: Asteroids that have remained largely unchanged since the early Solar System, composed of ancient materials that date back to the formation of the Solar System over 4.5 billion years ago. They are rich in carbon, water-bearing minerals, and organic compounds.

**Fully or partially differentiated asteroid**: Asteroids that have undergone internal differentiation, meaning its interior has separated into layers of different compositions due to heating and melting. This process typically occurs in larger asteroids once containing enough short-lived radioactive material (like $^{26}$Al) or experienced enough impact heating to cause partial or complete melting.

**Planetary embryo**: Large, solid celestial bodies that form during the early stages of planet formation. These embryos are larger than planetesimals but not yet fully developed planets, ranging in size from hundreds to thousands of kilometres in diameter.

**Non-carbonaceous meteorites (NC)**: A type of meteorite originates from asteroids formed in the inner Solar System and lacks significant amounts of carbon-rich material.

**Carbonaceous meteorites (CC)**: A type of meteorite that is rich in carbon, volatile elements, and organic compounds. Carbonaceous meteorites are fragments of primitive asteroids accreted in the outer Solar system.

*s*–**process nucleosynthesis**: Slow neutron-capture processes are nucleosynthetic pathways primarily occurring in asymptotic giant branch stars where atomic nuclei gradually capture neutrons over long timescales, allowing unstable isotopes to undergo beta decay before capturing additional neutrons. It is responsible for producing many of the heavy elements found in the Universe, such as strontium (Sr), barium (Ba), and lead (Pb).

***p*–process nucleosynthesis**: Proton capture or photodisintegration process are nucleosynthetic pathways responsible for producing proton-rich isotopes of heavy elements that cannot be formed by *s*– or *r*–processes. These isotopes are called *p*-nuclei and are found in elements such as molybdenum (Mo), ruthenium (Ru), and xenon (Xe), and is generally believed to occur supernovae environments.

***r*–process nucleosynthesis**: Rapid neutron-capture processes are nucleosynthetic pathways primarily occurring in neutron star mergers and core-collapse supernovae where atomic nuclei rapidly capture multiple neutrons before undergoing beta decay. This process is responsible for producing about half of the heavy elements beyond iron (Fe), including gold (Au), platinum (Pt), uranium (U), and thorium (Th).

**Viscous disk expansion**: Viscous expansion in a protoplanetary disk refers to the outward spread of gas and dust due to viscous forces within the disk. These forces arise from turbulence, magnetic fields, or other mechanisms that generate internal friction between disk particles.

**Nugget effect**: In isotope geochemistry, the nugget effect refers to the uneven distribution of isotopically distinct materials within a sample, leading to variability in measured isotope ratios.

**Hill sphere**: The Hill sphere of an astronomical body is the region around it where its gravitational influence dominates over that of a more massive body it orbits.